\documentclass{svjour3}                     
\smartqed  

\usepackage{natbib}

\usepackage{booktabs} 
\usepackage{listings}
\usepackage{xcolor}
\usepackage[normalem]{ulem}
\usepackage{multirow}
\usepackage{url}
\usepackage{tabularx}
\usepackage{framed}

\newcommand{\listingsfont}{\footnotesize\ttfamily}

\usepackage{graphicx}
\begin{document}

\title{A Comparative Study of Application-level Caching Recommendations at the Method Level}

\titlerunning{A Comparative Study of Application-level Caching}        

\author{Rômulo Meloca         \and
        Ingrid~Nunes 
}


\institute{R. Meloca and I. Nunes \at
              Instituto de Inform\'{a}tica,
              Universidade federal do Rio Grande do Sul\\
              Porto Alegre, Brazil \\ 
              \email{\{rmmeloca,ingridnunes\}@inf.ufrgs.br}           
}

\date{Received: date / Accepted: date}

\maketitle

\begin{abstract}
Performance and scalability requirements have a fundamental role in most large-scale software applications. To satisfy such requirements, caching is often used at various levels and infrastructure layers. Application-level caching---or memoization---is an increasingly used form of caching within the application boundaries, which consists of storing the results of computations in memory to avoid re-computing them. This is typically manually done by developers, who identify caching opportunities in the code and write additional code to manage the cache content. The task of identifying caching opportunities is a challenge because it requires the analysis of workloads and code locations where it is feasible and beneficial to cache objects. To aid developers in this task, there are  approaches that automatically identify cacheable methods. Although such approaches have been individually evaluated, their effectiveness has not been compared. We thus in this paper present an empirical evaluation to compare the method recommendations made by the two existing application-level caching approaches at the method level, namely APLCache and MemoizeIt, using seven open-source web applications. We analyse the recommendations made by each approach as well as the hits, misses and throughput achieved with their valid caching recommendations. Our results show that the effectiveness of both approaches largely depends on the specific application, the presence of invalid recommendations and additional configurations, such as the time-to-live. By inspecting the obtained results, we observed in which cases the recommendations of each approach fail and succeed, which allowed us to derive a set of seven lessons learned that give directions for future approaches to support developers in the adoption of this type of caching.
\end{abstract}


\section{Introduction}\label{sec:introduction}

Many software systems that rely on the Internet---such as web applications and those in the context of Internet of Things and cloud computing---commonly have to meet increasing non-functional requirements. These include reduced response time and the ability to deal with a high number of requests. To support a growing number of simultaneous requests, developers must find a means of  speeding up computations or improve the underlying hardware. Even though the scaling up of hardware is frequently an effective way to address performance and scalability requirements, it incurs in financial costs that can grow exponentially~\citep{abbott&fisher2009}. Consequently, developers must find alternatives to satisfy such requirements. For this purpose, caching is typically used at various levels and application infrastructure to reduce the response time. Although some forms of caching are transparent to application developers, such as by using a proxy server, a recent form of caching, namely \emph{application-level caching}, has been adopted in a complementary way to other forms of caching by exploiting specificities of individual applications~\citep{Mertz2017understanding}.

Application-level caching, or memoization, is the use of caching within the boundaries of a software application. It requires developers to identify opportunities where the result of an already computed value can be reused and to add additional code to manage cached objects. This type of caching can be employed, for instance, at the database layer~\citep{Scully2017}, client-side~\citep{Huang2010}, or within the business logic~\citep{DellaToffola2015,Mertz2018}. Wherever such caching opportunities reside, it is not a trivial task to find them because, not only spots where computed values that are eligible to be reused must be found, but also the application workload must be inspected to identify values that are frequently accessed or take a long time to compute. Moreover, as the workload evolves, caching opportunities must be constantly revised to avoid performance decay. Consequently, this is a time-consuming and error-prone task~\citep{Mertz2017qualitative}, and developers benefit from supporting solutions.

Different approaches have been proposed to help developers employ caching at the application level. Some of them focus on providing recommendations of cacheable spots~\citep{xu2012finding, xu2013resurrector, cachetor2013, Maplesden2015, DellaToffola2015, chen2018speedoo}, while others aim to automate the identification and management of cached content at this level~\citep{Mertz2018, Wang2014}. In many cases, these approaches are \emph{complementary} because they target different granularity levels---\emph{complex objects} have been examined by four approaches~\citep{xu2012finding, xu2013resurrector, Maplesden2015, chen2018speedoo}, while three works focus on identifying \emph{methods to be cached}~\citep{cachetor2013, DellaToffola2015, Mertz2018}.

Our work focuses in particular on approaches that identify cacheable methods because by knowing which methods to cache, developers can improve the application performance without major changes in the source code. \citet{cachetor2013} were the first to explore this performance opportunity. However, they examined only recurrent method outputs, without assessing neither whether the method is feasible to be cached (i.e.\ produce the same output for the same input) nor if it would provide benefits if cached. There are only two approaches that aim to do this, namely \emph{APLCache}~\citep{Mertz2018} and \emph{MemoizeIt}~\citep{DellaToffola2015}. Although they have been individually evaluated, there is a lack of studies that compare them.

We thus in this paper fulfil this gap and present an empirical study that compares APLCache and MemoizeIt. Our goal is to compare their suggestions of cacheable spots and evaluate their effectiveness in terms of hits, misses and throughput, using seven open-source projects of web applications written in Java. All selected applications include caching manually implemented by developers. Our study has a two-phase procedure. First, we collect execution traces of each application based on a ``learning'' workload. These traces are given as input for both application-level caching approaches. As result, we obtain the recommendations made by APLCache and MemorizeIt. Each recommendation is then inspected for validity. Second, we execute ``testing'' workloads with four versions of each application: (i) application with the caching made by developers; (ii) application with valid recommendations made by APLCache; (iii) application with valid recommendations made by MemoizeIt; and (iv) application with no caching. 

Our results show that both approaches are able to discover opportunities different from those made by the developers. However, suggested opportunities must be carefully inspected because they include methods that will lead to bugs if cached. Considering the valid recommendations, both approaches are able to identify opportunities that generate hits. With respect to hits, misses and throughput, we observed diverging results across the different applications as well as number of simultaneous users. By qualitatively analysing the collected data, we identified the strengths and weaknesses of each approach. This knowledge is the basis for further improvements to be made in both APLCache and MemoizeIt, or even for developing future approaches.

Our key contributions are the following: (i) a study protocol for evaluating and comparing application-level caching approaches at the method level; (ii) an empirical comparison between APLCache and MemoizeIt; and (iii) seven implications of our study that give directions for improving existing approaches and key aspects that remain unaddressed.

We next provide a background on application-level caching as well as overview work on this topic, providing details of the compared approaches. We describe in Section~\ref{section:study-design} our study design and details of the target web applications. In Section~\ref{section:results-and-analysis}, we present and analyse the results of our study. Lessons learned and threats to validity are discussed in Section~\ref{section:discussion}. Finally, conclusions are presented in Section~\ref{section:conclusions}.

\section{Background and Related Work}
\label{section:background-related-work}

Much work has been done to support developers to improve the performance of software applications. There are approaches that focus on code optimisation. They propose, for example, rearranging instructions in the source code to avoid the repeated instantiation of objects inside loops~\citep{xu2012static} or inefficient recurrent JavaScript instructions~\citep{Selakovic2016}. Other approaches aim to prioritise performance optimisation opportunities of a method~\citep{chen2018speedoo} or improve the way that profilers show provided results by better ranking and aggregating them~\citep{Maplesden2015, Maplesden2015a}. Our work focuses on the use of caching at the application level to improve the application performance. In this section, we introduce background on this topic and discuss supporting approaches (Section~\ref{section:application-level-caching}), and then provide details of the two approaches compared in our study (Section~\ref{section:compared-approaches}).

\subsection{Application-level Caching}
\label{section:application-level-caching}

Application-level caching~\citep{Mertz2017understanding} consists of the use of caching for storing the results of computations (such as of a method or a set of instructions) to be reused in the future. This is often manually implemented in the source code, such as in the example presented in Listing~\ref{listing:app-level-caching-example}. In this example, a \texttt{Map} is used to cache the result of the \texttt{getBalance()} method, which retrieves data from the database only if it is not cached. Alternatively to a local map, cache components can be used, such as Redis or Caffeine. In addition to deciding what to cache~\citep{Mertz2018}, developers must also deal with, e.g., the invalidation of cached content in case of changes~\citep{Ghandeharizadeh2012} and a definition of a time-to-live (TTL)~\citep{Alici2012}. Caching opportunities in the code are computations that result in the same output for the same input at least for a period of time, and do not involve statements that cannot be skipped (e.g.\ a writing operation). Caching these computations saves computation time, thus providing performance gains and helping increase scalability, if they have high cost to compute and/or are frequently executed. Manually finding and managing these caching opportunities is a challenge and, therefore, approaches have been proposed to support this. 

	\lstset{
	  emph={AccountsManager, getBalance},
	  emphstyle={\bf}
	}
\begin{lstlisting}[caption={Example of a caching at the application level.}, label=listing:app-level-caching-example, xleftmargin=0.04\textwidth, xrightmargin=0.01\textwidth, float, belowskip=-2em]
class AccountsManager {

  Map<Account, Double> cache
      = new HashMap<>();
  AccountsDAO accountsDAO
      = new AccountsDAO();

  Double getBalance(Account account) {
    return cache.
        computeIfAbsent(account, key
            -> accountsDAO.retrieveBalance());
  }
}
\end{lstlisting}
	
Because application-level caching is a cross-cutting concern, imposing maintainability challenges, there are approaches that help \emph{managing the cache component}. In this direction, there is work that (1) searches for the best way to optimise memory allocation~\citep{Radhakrishnan2004, Ghandeharizadeh2015}; (2) improves admission or replacement algorithms~\citep{Qin2014, Venketesh2009, AliAhmed2011, Yang2003, Saemundsson2014, Santhanakrishnan2006, Ali2012a}; (3) provides adaptive caching policies~\citep{Alici2012, Subramanian2006, Megiddo2004}; or (4) automatically manages the caching at the server side~\citep{Ports2010}. In addition, there are approaches~\citep{Xu2014, Zaidenberg2015, Wood2013} that focus on addressing issues of a specific cache component, namely Memcached.
Given that caching opportunities are often located at the application boundaries, there are approaches that focus on \emph{specific locations to cache}. Most of them deal with various issues associated with caching at the \emph{data layer}~\citep{Ghandeharizadeh2012, Sun2017, Leszczynski2010, Scully2017, Wang2014, Gupta2011, Chen2016, Larson2004}. However, there are  approaches that focus on caching content at the \emph{presentation layer}~\citep{Huang2010, Candan2001} or deal specifically with \emph{files}~\citep{Guo2011}.

With respect to the \emph{identification of caching opportunities}, there is work that targets \emph{data structures} and \emph{methods}. Approaches that focus on the former search for reusable data structures~\citep{xu2012finding,cachetor2013} or disposed objects~\citep{xu2013resurrector}, helping reduce the memory consumption. The work by Nguyen and Xu~\citep{cachetor2013} also has a component that target methods, but it is limited to the identification of methods that produce recurrent results (regardless of the input) and does not assess the benefit of caching. We refer to work in this direction as \emph{method-level approaches}~\citep{DellaToffola2015, Mertz2018}, which aim at identifying methods that result in outputs that are possible to cache (i.e.\ produce the same output for the same input for a certain time frame) and would provide performance improvements if cached. Our study focuses on comparing these method-level approaches, and they are further detailed in next section.

\subsection{Compared Method-level Approaches}
\label{section:compared-approaches}

From the existing application-level caching approaches introduced above, only two explore methods inputs and outputs to identify those that are possible to be cached and/or provide benefits if cached. These approaches are APLCache~\citep{Mertz2018} and MemoizeIt~\citep{DellaToffola2015}. Both of them rely on the runtime analysis of the application execution considering a particular workload. They involve algorithms, metrics, and heuristics to suggest cacheable methods. In the following sections, we provide an overview of how each approach works, describing the rationale for their suggestion in a high-level way. For a complete understanding of the details of each approach, we refer the reader to their original publications.

\subsubsection{APLCache} 

\citet{Mertz2018} proposed an approach to automate the use of application-level caching. It is implemented as a framework, named APLCache, to be seamlessly integrated to web applications. The framework monitors the application at runtime collecting execution traces, identifies cacheable spots, and manages content to be cached. The decision of what to cache is based on the Cacheability Pattern, which was derived from a qualitative study~\citep{Mertz2017qualitative} of application-level caching in web applications. The implementation of the pattern consists of the definition of five metrics---staticity, changeability, frequency, shareability, and expensiveness---that are used to select content to be cached. These metrics are used to select methods that, in addition to producing the same outputs for the same inputs for certain periods of time, have some of the following properties: (i) are frequent; (ii) their results are shared among multiple users; and (iii) are expensive to compute. Generally, these metrics are assessed for all methods, and the mean and standard deviation are used as the basis to make decisions. A particularity of this approach is that it does not recommend to cache all pairs of inputs and output of a selected method, but only those that satisfy the properties introduced above. To compare and store inputs and outputs, this approach serialises objects. APLCache has been evaluated using three web applications, using as baseline the original application (which includes the caching made by developers) and the application with no caching. The module of APLCache that chooses cacheable methods can be used as a recommender, which is considered in our study.

\subsubsection{MemoizeIt}

A recommender of cacheable methods, named MemoizeIt, has been proposed by \citet{DellaToffola2015}. Their approach consists of four steps: time and frequency profiling, input-output profiling, clustering and ranking, and suggest cache implementation. This first step discards methods that are called only once or are not expensive to compute (according to a given threshold). Then, the remaining candidates are refined, discarding every method that has any different output for the same input. To reduce the computation time of this step, it is possible to execute it either in the exhaustive or the iterative mode. The former compares the whole object reference tree, while the latter partially compares inputs and outputs, truncating them at some point. The result of this step is a set of methods that are feasible to be cached, which are ranked according to their potential saved computation time. The third step aggregates methods that are called in a sequence, based on the analysis of the call graph of the application. Finally, the last step suggests how to implement the caching of objects. The implementation can vary in \emph{size}, storing a single or multiple instances of the results of a method, and in \emph{scope}, which can be an instance or global (static) map. MemoizeIt was evaluated with eleven single-threaded programs.

APLCache and MemoizeIt have two key differences. First, when MemoizeIt points out a method to be cached, any pair of inputs-output is cached if invoked. In contrast, APLCache selects specific pairs to be cached. Second, APLCache admits caching the result of a computation based on a set of inputs even if the results vary in a certain time frame.
In certain applications, stale data are allowed for a short period of time in these cases, or the cached object is invalidated.
MemoizeIt, in turn, only recommends methods that have invariant results for a set of inputs. Note that both approaches consider only the method inputs and output, not analysing if there are instructions within the method that cannot be skipped. MemoizeIt is used as a recommender and thus suggestions must be manually inspected. APLCache, in turn, acknowledges this problem and suggests that developers annotate uncacheable methods that fall in this situation. However, this requires manually analysing all application methods, which is a time-consuming and error-prone activity.

\section{Study Settings}
\label{section:study-design}

Given that we introduced the two approaches that are compared in our work, we now proceed to the description of our study settings.

\subsection{Research Questions}

The goal of our work is to evaluate and compare method recommendations made by two application-level caching approaches. In order to achieve this goal, we focus on answering two research questions, detailed as follows.

    \begin{itemize}
        \item \textbf{RQ1}: What are the differences between the recommendations made by APLCache and MemoizeIt?
        \item \textbf{RQ2}: What are the performance improvements provided by the valid caching recommendations of each approach?
    \end{itemize}
    
With RQ1, we aim to analyse the recommendations made by each approach and compare them with each other and the caching decisions made by developers. With RQ2, we focus on assessing the improvements provided by the caching recommendations in term of throughput, hits and misses, which are the typical metrics used to evaluate caching. We use as baseline applications with no caching and the caching implemented by developers. Hereafter, we refer to APLCache, MemoizeIt, and the developers' implementation as APL, MEM, and DEV, respectively.

\subsection{Procedure}
\label{section:procedure}

\begin{figure}
    \centering
    \includegraphics[width=.6\linewidth]{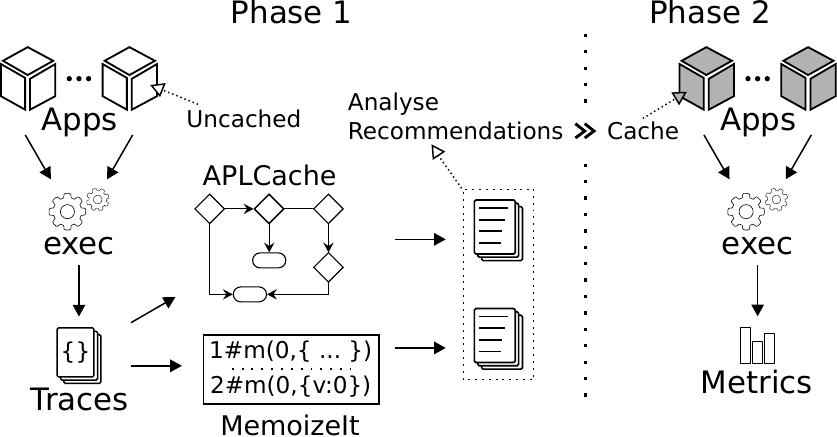}
    \caption{Overview of the study procedure.}
    \label{figure:study-design}
\end{figure}

To answer our research questions, we specified a \emph{two-phase} procedure, which is overviewed in Figure~\ref{figure:study-design}. In phase 1, we first generate a synthetic workload in a systematic way and execute a set of web applications with no caching and collect their execution traces. Not only is a real workload  typically not freely available, but also it would not provide benefits to our study. The goal is to assess the ability of application-level caching approaches to learn usage patterns. Consequently, they should be able to learn any typical usage pattern. The collected execution traces are then provided as input for APL and MEM to generate caching recommendations.\footnote{Note that if new application-level caching approaches are proposed to recommend methods to cache, the same procedure can be used. The only requirement is to use the recommendations made by the new approach. We provide details on how to reproduce the study in Appendix~\ref{appendix:reproducibility} for those who would like to use our infrastructure to conduct similar studies.} As they can be invalid, we inspect them discarding those that cannot be cached. Next, in phase 2, we implement three versions of the application, caching methods based on APL, MEM, and DEV. These versions as well as the application with no caching are executed using another synthetic workload, but following the same set of probabilities of navigation of the previous workload so that there is a usage pattern. Finally, we collect metrics to be analysed.\footnote{The source code of our study is available for reproducibility at \url{http://inf.ufrgs.br/prosoft/resources/2021/emse-apl-caching-comparison}.} We next provide details of each step of our study.

\subsubsection{Phase 1}

\textbf{Application Adaptation and Workload Generation.} We selected a set of target applications (described in next section) that contain on its implementation cached methods. These applications were modified by removing all caching snippets, creating their uncached version, which we refer to as NOCACHE. To be able to execute the applications, we automatically generated data for those that do not have a database dump available. Moreover, we manually inspected each application and generated a navigation structure stored in a JSON file. This file contains the information of two graphs $G_{N} = (V, E_{N})$, which gives the possible HTTP requests that can be done after another, and $G_{R} = (V, E_{R})$, which gives the HTTP requests that are required to be executed before another. Each vertex $v \in V$ represents a HTTP request. Each $(v, u) \in E_{N}$ indicates that the HTTP request $u$ is allowed to be subsequently invoked after the HTTP request $v$ according to the business rules of the application. Each $(v, u) \in E_{R}$ indicates that the HTTP request $v$ must have been executed before the HTTP request $u$ is invoked. For example, a product can be updated only after being visualised. The execution of $v$ guarantees that, e.g., required attributes are stored in the local session of the application. A synthetic workload is then generated by selecting a sequence of requests that are in accordance with the navigation structure in the JSON file. A possible next request is randomly selected based on the previous requests. This selection also considers the proportion of 80\% of read-only requests (GET) and 20\% of write requests (POST, PUT and DELETE), which is in accordance with the TPCW\footnote{\url{http://www.tpc.org/tpcw}} benchmark and used in previous work~\citep{Mertz2018}. The generation of each request contains the required headers and form data.

\textbf{Collection of Execution Traces.} To have a set of execution traces, we use the default Java implementation for basic HTTP requests and Apache implementation to multipart requests in our simulator. We sent requests for five minutes for each application. We used a dedicated machine to simulate the requests (Intel Core i5 processor and 8 Gb of main memory) and another to host the applications (Intel Core i7 processor and 32 Gb of main memory). Using AspectJ\footnote{\url{https://www.eclipse.org/aspectj}}, we intercepted every method call, recording it in a file with JSON objects. Each trace contains the time spent to compute (start and end times), as well as the HTTP user session, the method signature, input parameters and the produced result. 
In order to serialise parameters and the return of methods to JSON, we adapted JSON-java\footnote{\url{https://github.com/stleary/JSON-java}} to break cycles and to truncate packages not related to the application. We serialised each field of objects, recursively, instead of its getters. To decide whether a field should be truncated during the serialisation we consulted the POM file searching for a dependency that matches the package name of the field. Fields belonging to internal Java packages are converted to a string (\texttt{String.valueOf()}) instead of being recursively explored.

\textbf{Recommendation Generation and Analysis.} APL and MEM had to be adapted to generate caching recommendations using our traces.
For APL, we had to extract from its framework the component that decides what to cache. For MEM, we implemented it using the original implementation as a basis, which could not be used because it consists of an instrumentation in the source code to collect information at runtime. We, instead, provide as input a previously generated set of traces that had to be processed. Our implementation includes both the iterative and exhaustive kernels. Moreover, MEM requires a callgraph of the application, and we use java-callgraph\footnote{\url{https://github.com/gousiosg/java-callgraph}} for this purpose. For deserialising traces, we use Gson\footnote{\url{https://github.com/google/gson}}. APL and MEM include a set of thresholds, for which we use the same values adopted by their authors. These thresholds are not required to be tuned for specific applications, They are general setups to give semantics for certain concepts of the approaches. For example, APL classifies methods as frequent when they are k standard deviations above the average. MEM indicates with a threshold when a method executes so fast that it should not even be considered for caching. We thus used the threshold configurations proposed by the authors. By running the compared approaches, we collected their respective sets of recommendations. Each recommendation was manually analysed---taking into account the method and its context---to verify if it does not contain any command that must always be executed when the method is invoked. That is, if the method is cached, the application would result in a failure. Common circumstances in which a  recommendation is invalid are when the recommended method: (a) performs a writing operation in the database; (b) makes a request or sends a message to an external service; (c) writes into a file; (d) changes data stored in static fields; or (e) directly or indirectly manipulates the parameters given in its invocation. This verification was made by a careful manual analysis of the applications, complemented by the use of available unit tests. Recommendations that break unit tests are also considered invalid. 

\subsubsection{Phase 2}

\textbf{Caching Implementation.} Using Caffeine\footnote{\url{https://github.com/ben-manes/caffeine}} as a cache component, we created three versions of the application in addition to its uncached version. We implemented an APL and a MEM versions, using the valid recommendations of each approach. While for MEM every call to a cached method is searched in the cache component or stored in it, for APL only calls with particular inputs are cached. Therefore, before storing an entry in the cache component, we must verify if the corresponding input is an APL recommendation. We also replaced the original caching done by developers in the DEV version by using a standardised cache component and to be able to collect metrics that serve as baseline. Despite this change, the developers' caching decisions remained the same. Finally, for all cached versions, we configured the cache component with no size limit, in order to avoid influencing the results by the choice of a cache replacement policy that would automatically evict objects. Having a particular size limit could cause misses that are not due to choice of cached methods, favouring one of the approaches. Cached objects leave the cache when their TTL expires. We adopted the TTL values used by developers.

\textbf{Execution and Metrics Collection.} In order to measure the performance of each version of the application (NOCACHE, APL, MEM, and DEV) under the same conditions, we generated the workloads for each application based on their NOCACHE version and we ran it ten times on every version. In the generation the allowed random values and the chances of performing read or write operations---as discussed in Phase 1---requests are chosen and logged to be exactly the same for the next ten executions of each application version and number of simulated users.We simulated 1, 5, and 25 simultaneous users---similarly to previous work~\citep{Mertz2018}---firing requests during 10 minutes. In order to validate the number of users that should be simulated and duration of the simulation, we ran previous tests and observed no large differences in the behaviour of the applications and the cache component when simulating 5, 10, or 50 users. Moreover, we also found in previous testings that there is no need to collect data from 10 to 25 minutes of execution because there is also no behavioural change in such a period. This occurs because the workload follows a particular overall usage pattern (as it is expected in web applications that aim at exploiting application-level caching) and the caching recommendations are generated for that particular pattern. These workloads were then used to execute each version of each application 10 times to measure the \emph{throughput} (requests handled per second). An additional execution was performed to assess cache performance metrics. 
The cache performance metrics are mainly \emph{hits} and \emph{misses}, from which hit-ratio can be derived. Moreover, as APL does not necessarily put an object in the cache each time that there is miss (because it caches only particular inputs), we also collected the number of cache \emph{additions} and \emph{discarded} inputs.

Based on the caching recommendations and the collected metrics, we compare APL and MEM. Note that, although DEV is used as a baseline, the caching decisions made by developers cannot be directly compared with APL and MEM because developers made decisions based on their production environment. This is not an issue to compare APL and MEM because they generate recommendations for a given workload, regardless if it is synthetic or collected from a production environment. Consequently, DEV is used in our study as a reference and the developers' decisions are not in-depth discussed.

\subsection{Target Applications}
\label{section:target-applications}

For selecting our target web applications, we retrieved from GitHub the 1,000 most popular Java repositories according to stars. From these, we discarded those that do not contain a POM file that indicate at least one dependency to one of the common caching providers, namely Spring\footnote{\url{https://spring.io}}, Redis\footnote{\url{https://redis.io}}, Ehcache\footnote{\url{https://www.ehcache.org}}, Memcached\footnote{\url{https://memcached.org}} and Caffeine\footnote{\url{https://github.com/ben-manes/caffeine}}. We manually analysed the description of the resulting 492 repositories, from which 415 were excluded because they consist of example applications, are not web applications, or do not use English in the application source code. The remaining 77 repositories were manually inspected to select those that (i) include application-level caching already implemented by their developers through mappings or one of the mentioned tools, (ii) handle HTTP POST, and (iii) are feasible to adapt and create the versions needed for our study. We checked if: (1) the application does not involve mixed programming languages because we need to intercept the code using aspect orientation; (2) the caching was not implemented in an interleaved way with the business logic because changing such a code could lead to bugs while preparing the setup of our study; and (3) the applications provides a REST interface allowing us to make GET and POST HTTP requests. This resulted in nine repositories, from which we discarded two due to performance and non-reproducibility issues. 


The seven web applications that are used in our study are presented in Table~\ref{table:target-applications}, along with a description of their domain, numbers of lines of code (LOC) and number of files. As can be seen, they are from various domains and sizes. In Table~\ref{table:database-instance}, we present the number of cached opportunities manually implemented by the developers with dedicated code (as shown in Listing~\ref{listing:app-level-caching-example}) in each application as well as the amount of data generated to populate their databases. Azkaban have demo workflows available, while Thingsboard includes accounts with built-in data, which were used. Data entities presented in Table~\ref{table:database-instance} also have many relationships with other entities not listed in table.

\begin{table}
    \caption{Target applications.
    \label{table:target-applications}}
    \centering
    \begin{tabularx}{\linewidth}{l X r r r}
        \toprule
        \textbf{Application} & \textbf{Domain} & \textbf{LOC} & \textbf{\#Files} & \textbf{GitHub Stars} \\ 
        \midrule
        Azkaban 3.70.1 & A workflow manager & 77,484 & 473 & 2,125 \\
        Cloudstore 2.0.0 & An e-commerce application based on TPC-W benchmark & 11,228 & 98 & 16 \\
        Keycloak 5.0.0 & A manager for identity and access & 395,388 & 3,257 & 2,997 \\
        Killbill 0.20.9 & A subscription billing and payment platform & 134,279 & 1,018 & 1,788 \\
        Petclinic 1.5.0 & A sample Spring-based application & 1,566 & 25 & 2,704 \\
        Shopizer 2.2.0 & A customisable e-commerce application & 95,363 & 797 & 2,105 \\
        Thingsboard 2.3.0 & An IoT platform & 74,180 & 848 & 2,182 \\
        \bottomrule
    \end{tabularx}
\end{table}

\begin{table}
    \caption{Target applications: caching and database data.
    \label{table:database-instance}}
    \centering
    \begin{tabular}{l r l l}
        \toprule
        \textbf{Application} & \textbf{DEV Cached Methods} & \textbf{Input Data} & \textbf{Synthetic Data} \\
        \midrule
        Azkaban     & 4 & 2 demo workflows & No \\
        Cloudstore  & 4 & 10,000 items & Yes \\
        Keycloak    & 12 &  100 clients & Yes \\
        Killbill    & 30 & 100 accounts & Yes \\
        Petclinic   & 1 & 1,000 vets & Yes \\
        Shopizer    & 16 & 10,000 products & Yes \\
        Thingsboard & 13 & 3 demo clients & No \\ 
        \bottomrule
    \end{tabular}
\end{table}

\section{Results and Analysis}
\label{section:results-and-analysis}

By following the described procedure, we obtained the results presented in this section. Additional information about the data collected in the study together with complementary charts is available online.\footnote{\url{http://inf.ufrgs.br/prosoft/resources/2021/emse-apl-caching-comparison}}

\subsection{RQ1: Caching Decisions}
\label{section:rq-1}

The number of method recommendations made by APL and MEM for each application are shown in Table~\ref{table:recommendations}. The recommendations can be: (i) \emph{novel}, if the method was not cached by developers; (ii) \emph{existing}, if the method was cached by developers; or (iii) \emph{invalid}, if the recommendation corresponds to an uncacheable method. We also present how many of the valid recommendations are \emph{useful}, i.e.\ cached methods that led to hits, and the \emph{usefulness rate}, which is the percentage of valid recommendations that were useful.


\begin{table}
\renewcommand{\tabcolsep}{0.15cm}
\caption{Analysis of caching recommendations.}
\label{table:recommendations}
\centering
\begin{tabular}{ll | rrr | rr | r}
\toprule
\multicolumn{1}{c}{\multirow{2}{*}{}} &
  \multicolumn{1}{c}{\multirow{2}{*}{}} &
  \multicolumn{6}{c}{\textbf{Recommendations}} \\ \cmidrule(l){3-8} 
\multicolumn{1}{c}{\rotatebox[origin=l]{90}{\textbf{Application}}} &
  \multicolumn{1}{c|}{\rotatebox[origin=l]{90}{\textbf{Approach}}} &
  \multicolumn{1}{c}{\rotatebox[origin=l]{90}{\textbf{Novel}}} &
  \multicolumn{1}{c}{\rotatebox[origin=l]{90}{\textbf{Existing}}} &
  \multicolumn{1}{c|}{\rotatebox[origin=l]{90}{\textbf{Invalid}}} &
  \multicolumn{1}{c}{\rotatebox[origin=l]{90}{\textbf{Useful}}} &
  \multicolumn{1}{c|}{\rotatebox[origin=l]{90}{\textbf{Usefulness rate}}} &
  \multicolumn{1}{c}{\rotatebox[origin=l]{90}{\textbf{Total}}} \\
\midrule
\multirow{2}{*}{\rotatebox[origin=c]{0}{Azkaban}}  & APL & 6    & 0    & 0    & 6    & 1.00 & 6     \\
                                                  & MEM & 2    & 0    & 0    & 2    & 1.00 & 2     \\   \cmidrule(lr){3-8}
\multirow{2}{*}{\rotatebox[origin=c]{0}{Cloudstore}}  & APL & 3    & 0    & 3    & 3    & 1.00 & 6   \\
                                                  & MEM & 2    & 0    & 5    & 2    & 1.00 & 7     \\   \cmidrule(lr){3-8}
\multirow{2}{*}{\rotatebox[origin=c]{0}{Keycloak}}  & APL & 2    & 0    & 0    & 2    & 1.00 & 2     \\
                                                  & MEM & 7    & 0    & 5    & 5    & 0.71 & 12    \\   \cmidrule(lr){3-8}
\multirow{2}{*}{\rotatebox[origin=c]{0}{Killbill}}  & APL & 17   & 3    & 13   & 19   & 0.95 & 33    \\
                                                  & MEM & 25   & 3    & 40   & 24   & 0.86 & 68    \\    \cmidrule(lr){3-8}
\multirow{2}{*}{\rotatebox[origin=c]{0}{Petclinic}}  & APL & 3    & 0    & 2    & 3    & 1.00 & 5     \\
                                                  & MEM & 1    & 0    & 1    & 1    & 1.00 & 2     \\    \cmidrule(lr){3-8}
\multirow{2}{*}{\rotatebox[origin=c]{0}{Shopizer}}  & APL & 10   & 2    & 4    & 12   & 1.00 & 16    \\
                                                  & MEM & 24   & 2    & 10   & 22   & 0.85 & 36    \\    \cmidrule(lr){3-8}
\multirow{2}{*}{\rotatebox[origin=c]{0}{Thingsboard}}  & APL & 2    & 0    & 6    & 1    & 0.50 & 8     \\
                                                  & MEM & 2    & 0    & 1    & 2    & 1.00 & 3     \\
\midrule
\multirow{2}{*}{\rotatebox[origin=c]{0}{\textbf{Average}}} & APL & 6.14 & 0.71 & 4.00 & 6.57 & 0.92 & 10.86 \\
                                                  & MEM & 9.00 & 0.71 & 8.86 & 8.29 & 0.92 & 18.57 \\
\bottomrule
\end{tabular}
\end{table}

As can be seen, for most of the applications, both APL and MEM provide various invalid recommendations.
Their main criteria for considering a method cacheable is the analysis of method inputs and the corresponding output, and the content of the method is not taken into account. This is a crucial issue to provide better recommendations. For Killbill, for example, the number of MEM invalid recommendations is even higher than the valid recommendations. By analysing these invalid MEM suggestions, we observed that they mostly consist of request-mappings \textbf{Evidence 1}). These methods provide as output a constant but cannot be cached because of internal operations that cannot be skipped. Other recommended uncacheable methods are those that return an iterator instead of a list. To illustrate an invalid MEM recommendation, we show as follows a method of Petclinic, which would cause a failure while listing veterinarians to users if cached.
\begin{verbatim}
@GetMapping("/vets.html")
public String showVetList(Map<String, Object> model) {
  Vets vets = new Vets();
  vets.getVetList().addAll(findAll());
  model.put("vets", vets);
  return "vets/vetList";
}
\end{verbatim}

Invalid recommendations of APL, in contrast, consists mainly of methods that return the result of a computation but also include writing operations that must be carried out. Moreover, the APL invalid recommendations also include methods that are not intended to always result in the same output given the same input. For instance, APL recommended to cache the following method of Cloudstore, which based on the APL calculations was not considered changeable enough to be discarded.
\begin{verbatim}
public List<IItem> getPromotional() {
  Random rand = new Random();
  int randomId = rand.nextInt(1000) + 1;
  String hql = "SELECT * FROM item where item.id = " + randomId;
  . . .
  return items;
}
\end{verbatim}
This shows the importance of indicating to APL uncacheable methods because, otherwise, it would introduce bugs in the target applications if used as a self-adaptive approach, as its authors proposed. In addition, identifying uncacheable methods might be actually helpful for APL because, if these are discarded, the mean and standard deviations of the metrics used to make caching decisions would change, leading to another set of recommendations (\textbf{Evidence 2}).

In some situations, the number of APL and MEM recommendations is similar. However, by inspecting the recommended methods, there is little intersection between them. This suggests that these two application-level caching approaches are complementary. Not only are they different from each other, but they also recommend methods that were not cached by the developers and, as shown by the usefulness rate, these methods generate hits, potentially providing performance improvements. 

A main factor that justifies the differences between the two approaches is that APL admits \emph{some} degree of changeability in the cached methods. Although this can cause stale data in the application, this is accepted by developers when caching other methods in the applications, relying only on the TTL to invalidate cached content. Developers must thus be careful to verify if undesired stale data is caused by caching a recommended method.\footnote{Note that deciding the validity of the recommendation is not possible in these cases without discussing the requirements of each application with involved stakeholders. We considered these valid recommendations.} Moreover, MEM cannot guarantee that the recommended methods never change because it relies solely on the observed executions. Nevertheless, it is less likely to recommend methods that cause stale data. Consequently, when analysing suggested recommendations, developers should consider false positives for APL (recommended methods that could cause undesired stale data) and false negatives for MEM (not recommended methods that, although changeable, can be cached and provide benefits if so). This explanation justifies methods recommended by APL but not by MEM. Recommendations made only by MEM occur because APL assumes that methods that are not frequent, expensive, or shared by multiple users would not provide benefits if cached, even if they are not changeable (\textbf{Evidence 3}). Therefore, it does not recommend methods in this case, while MEM does.

As mentioned in our study procedure, DEV is used as a reference in our study. However, we observed a main general difference between both APL and MEM and the developers' choices is that developers, in some occasions, cached only lower-level methods, while the compared recommenders suggested higher-level methods (\textbf{Evidence 4}). For example, in the Petclinic application, the developers cached the method \texttt{Collection<Vet> findAll()}, while APL and MEM recommended to cache a method that calls the above method, shown below.
\begin{verbatim}
Vets showResourcesVetList() {
  return new Vets().getVetList().addAll(
    findAll()
  );
}
\end{verbatim}
As can be seen, the developers cached only the database access. However, APL and MEM also prevented the re-instantiation of the \texttt{Vets} data structure. Of course, for this to be possible, the code must be inspected to know if this data structure can indeed be reused (which is the case).

Now, we look into the particularities of each approach. APL recommends to cache particular inputs of a method, not all of them, assuming that only a subset of the inputs would promote benefits if cached. In Table~\ref{table:inputs-discarded}, we provide data associated with such inputs. By comparing the Cached column of this table (which shows the number of distinguished cached inputs) and the Novel and Existing columns of Table~\ref{table:recommendations}, it is possible to see that only a few inputs per method are cached (\textbf{Evidence 5}). For example, Keycloak has two Novel or Existing cached methods as well as two cached inputs. Consequently, one input was cached from each method. Petclinic, in turn, had three Novel or Existing cached methods, but 12 cached inputs. This means that at least one method had more than one input cached. In the second phase of our study, we measured---from cached methods---how many calls occurred with uncached inputs and, from these, how many are distinct. These data are shown in Table~\ref{table:recommendations}, for 1, 5 and 25 users. We observed that many inputs are recurrent and therefore could be cached. Nevertheless, APL can only recommend inputs that were observed in traces used to produce recommendations. Consequently, methods that have recurrent inputs that vary over time are not adequately cached. The complete APL solution focuses on constantly monitoring the application and updating the cached inputs and methods. Although this mechanism addresses this input issue, the solution must be used with caution to not create bugs by automatically caching invalid recommendations.


\begin{table}
\renewcommand{\tabcolsep}{0.15cm}
\caption{Inputs discarded by APLCache.}
\label{table:inputs-discarded}
\centering
\begin{tabular}{l | r | rrr | rrr}
\toprule
  \multicolumn{1}{c|}{\multirow{3}{*}{\textbf{Application}}} &
  \multicolumn{1}{c|}{\multirow{3}{*}{\textbf{Cached}}} &
  \multicolumn{6}{c}{\textbf{Uncached Inputs}} \\
\multicolumn{1}{c|}{} &
  \multicolumn{1}{c|}{} &
  \multicolumn{3}{c|}{\textbf{Distinct}} &
  \multicolumn{3}{c}{\textbf{Occurrences}} \\
\multicolumn{1}{c|}{} &
  \multicolumn{1}{c|}{} &
  \multicolumn{1}{c}{\textbf{1}} &
  \multicolumn{1}{c}{\textbf{5}} &
  \multicolumn{1}{c|}{\textbf{25}} &
  \multicolumn{1}{c}{\textbf{1}} &
  \multicolumn{1}{c}{\textbf{5}} &
  \multicolumn{1}{c}{\textbf{25}} \\ \midrule
Azkaban & 8  & 2,769  &  8,836  &  10,730  &  645,950 & 2,177,048 & 2,579,975  \\
Cloudstore & 2  &   600  &    600  &     600  &    3,962 &    22,494 &    56,293  \\
Keycloak & 2  &     1  &      1  &       1  &    4,735 &    24,748 &    36,657  \\
Killbill & 25 &    15  &     15  &      15  &    2,742 &    17,084 &    25,235  \\
Petclinic & 12 &    68  &     84  &      85  &    1,772 &     6,589 &     6,468  \\
Shopizer & 12 &    12  &     12  &      12  &   50,233 &   301,519 & 1,243,368  \\
Thingsboard & 3  &     5  &      5  &       5  &       30 &        30 &        30  \\
\bottomrule
\end{tabular}
\end{table}

Differently from APL, MEM not only recommends cacheable methods but also how to implement them, as explained in Section~\ref{section:background-related-work}. We detail the suggested implementation alternatives in Table~\ref{table:memoizeit-recommendation-type}. For all valid recommendations, MEM recommended the use of a global cache, shared among the various instances of a class. With respect to the size of the cached used for a method, most of the recommendations consist of single-instance caches. This means that for a particular method, there is only one entry in the cache and, if an input that is not in the cache is provided as parameter of a method call, the method is executed and its output is cached replacing the previous cached entry (\textbf{Evidence 6}). In only six cases (of the Killbill application), the cache can hold multiple entries in the cache for a method. On the one hand, single-instance caching prevents overpopulating the cache, as each method is limited to a single entry in the cache. On the other hand, if the input of a cached method with single-instance caching begins to vary, the method is always executed and its output replaces the cached object, leading to a performance worse than if the method is not cached. This type of implementation should, therefore, be used with a solid understanding of the method behaviour. MEM does not analyse the range of possible inputs (based on traces) of methods, but evaluates considering the observed traces which implementation leads to better performance results. Lastly, there are few cached getter methods, which are those that have no input. In these cases, the result is constant and the method is only re-executed if the TTL of the cached entry expires.


\begin{table}
\renewcommand{\tabcolsep}{0.15cm}
\caption{Types of valid MEM recommendations.}
\label{table:memoizeit-recommendation-type}
\centering
\begin{tabular}{l | rr | rrr}
\toprule
\textbf{Application} & \textbf{Instance} & \textbf{Global} & \textbf{Getter} & \textbf{Single-instance} & \textbf{Multi-instance} \\
\midrule
\multirow{1}{*}{\rotatebox[origin=c]{0}{Azkaban}} & 0 & 2 & 0 & 2 & 0 \\
\multirow{1}{*}{\rotatebox[origin=c]{0}{Cloudstore}} & 0 & 2 & 0 & 2 & 0 \\
\multirow{1}{*}{\rotatebox[origin=c]{0}{Keycloak}} & 0 & 7 & 3 & 4 & 0 \\
\multirow{1}{*}{\rotatebox[origin=c]{0}{Killbill}} & 0 & 27 & 0 & 21 & 6 \\
\multirow{1}{*}{\rotatebox[origin=c]{0}{Petclinic}} & 0 & 1 & 1 & 0 & 0 \\
\multirow{1}{*}{\rotatebox[origin=c]{0}{Shopizer}} & 0 & 30 & 2 & 28 & 0 \\
\multirow{1}{*}{\rotatebox[origin=c]{0}{Thingsboard}} & 0 & 2 & 0 & 2 & 0 \\ 
\bottomrule
\end{tabular}
\end{table}

MEM has another particularity, which is the choice for two alternative kernels for generating recommendations. The exhaustive kernel provides the complete results (which are those reported in the paper), while the iterative kernel truncates the analysis to gain performance. This caused one method of Shopizer to be erroneously recommended as cacheable by the iterative kernel. In addition, this kernel also discards two methods of Killbill that could be cached. This occurred because two inputs are considered equal (based on analysis of part of the their object trees) when they are not (based on their complete object trees) but the outputs are different. Therefore, the iterative kernel erroneously discarded the method because the supposed same input generated different outputs. Although the exhaustive kernel recommends them, they are invalid recommendations due to internal method operations.


\subsection{RQ2: Performance Measurements}

Having analysed the recommendations made by APL and MEM, we now examine the impact that caching these recommendations have on the performance of the applications.\footnote{We remind the reader that additional charts to further inspect the results are available at \url{http://inf.ufrgs.br/prosoft/resources/2021/emse-apl-caching-comparison}.} The results are shown in Table~\ref{table:throughput}, which shows how much each set of cached methods (DEV, APL, and MEM) increases or decreases the throughput (requests per second) of the applications in relation to the NOCACHE version. We present the median of the set of our ten executions---the higher the median, the better. Additionally, Table~\ref{table:throughput} also presents the difference between the first and third quartiles, which indicates the variance among the executions. The lower the difference, the lower the variance.

\begin{figure}
    \centering
    \includegraphics[scale=0.85,angle=90]{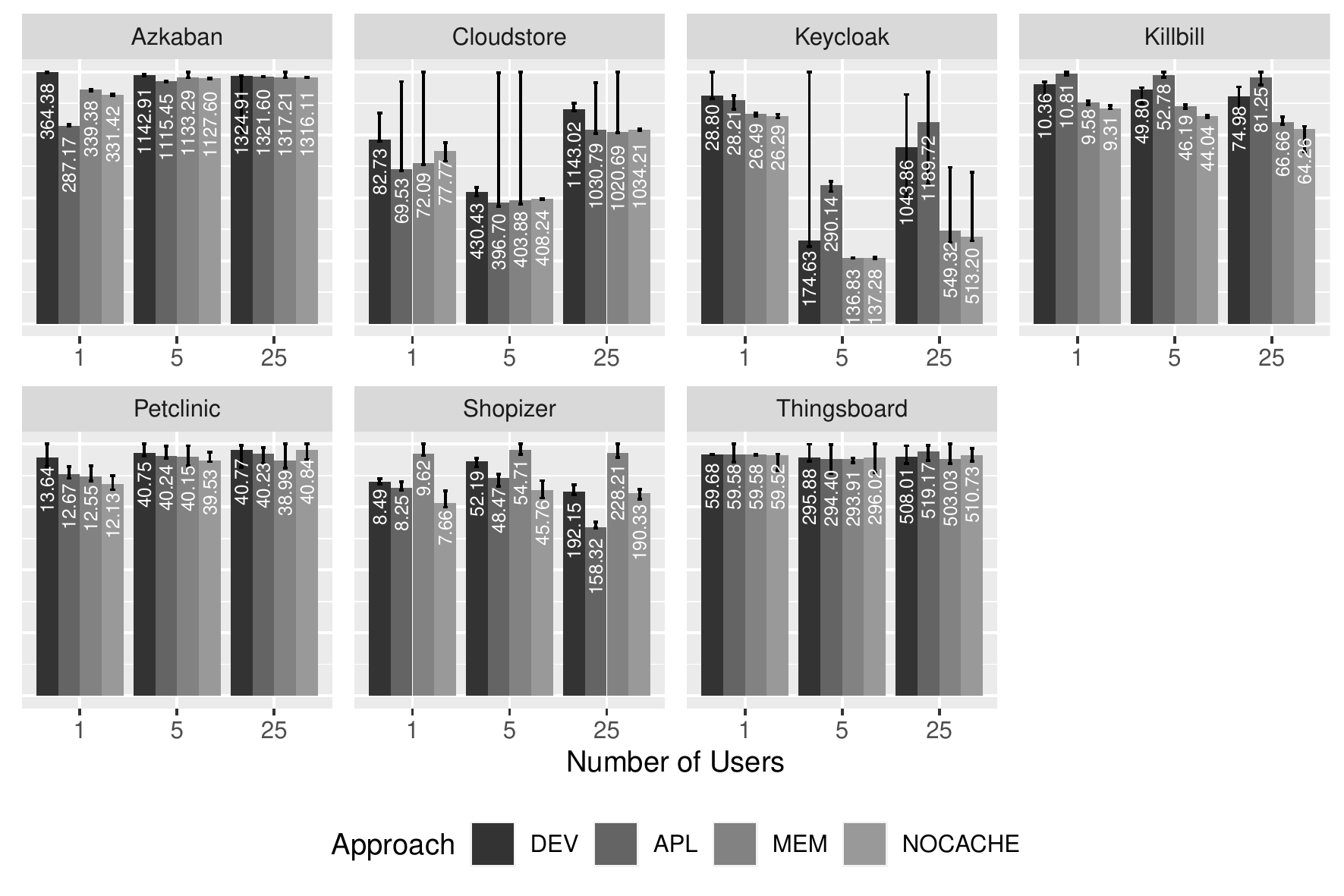}
    \caption{Normalised throughput (requests per second) by approach for each application.}
    \label{figure:throughput}
\end{figure}

Figure~\ref{figure:throughput} complements this table by presenting a bar chart of the normalised throughputs (with error) labelled with the actual throughput. The normalisation has the purpose of easing the comparison across the different applications. For two applications (Cloudstore and Keycloak) one or two executions are outliers because some requests resulted in crashes (which also occur in the original applications), leading to higher throughputs. Given that we consider the median value, the results are not disturbed by these outliers.

\begin{table}[]
\renewcommand{\arraystretch}{1.35}
\centering
\caption{Throughput compared to the \emph{NOCACHE} version and Quartile Increase.}
\label{table:throughput}
\resizebox{\textwidth}{!}{%
\begin{tabular}{@{}ll | rrrrrr@{}}
\toprule
\multicolumn{1}{c}{\multirow{3}{*}{\textbf{Application}}} &
  \multicolumn{1}{c}{\multirow{3}{*}{\textbf{Approach}}} &
  \multicolumn{6}{c}{\textbf{Relative Throughput and Quartile Increase}} \\
\multicolumn{1}{c}{} &
  \multicolumn{1}{c}{} &
  \multicolumn{2}{c}{\textbf{1 User}} &
  \multicolumn{2}{c}{\textbf{5 Users}} &
  \multicolumn{2}{c}{\textbf{25 Users}} \\
\multicolumn{1}{c}{} &
  \multicolumn{1}{c}{} &
  \multicolumn{1}{c}{\textbf{Median}} &
  \multicolumn{1}{c}{\textbf{Q. Inc.}} &
  \multicolumn{1}{c}{\textbf{Median}} &
  \multicolumn{1}{c}{\textbf{Q. Inc.}} &
  \multicolumn{1}{c}{\textbf{Median}} &
  \multicolumn{1}{c}{\textbf{Q. Inc.}} \\ \midrule
\multirow{3}{*}{Azkaban}          & DEV & 9.94\%   & 0.18\% & 1.36\%   & 0.13\%   & 0.67\%   & 0.17\%  \\
                                  & APL & -13.35\% & 0.40\% & -1.08\%  & 0.46\%   & 0.42\%   & 0.13\%  \\
                                  & MEM & 2.40\%   & 0.30\% & 0.51\%   & 0.13\%   & 0.08\%   & 0.12\%  \\ \hline
\multirow{3}{*}{Cloudstore}       & DEV & 6.39\%   & 0.68\% & 5.43\%   & 1.41\%   & 10.52\%  & 0.50\%  \\
                                  & APL & -10.59\% & 0.99\% & -2.83\%  & 1.18\%   & -0.33\%  & 0.66\%  \\
                                  & MEM & -7.29\%  & 1.51\% & -1.07\%  & 0.58\%   & -1.31\%  & 0.70\%  \\ \hline
\multirow{3}{*}{Keycloak}         & DEV & 9.56\%   & 3.06\% & 27.21\%  & 171.80\% & 103.40\% & 41.33\% \\
                                  & APL & 7.32\%   & 2.65\% & 111.34\% & 2.75\%   & 131.82\% & 29.60\% \\
                                  & MEM & 0.75\%   & 1.05\% & -0.33\%  & 0.91\%   & 7.04\%   & 29.12\% \\ \hline
\multirow{3}{*}{Killbill}         & DEV & 11.24\%  & 0.56\% & 13.08\%  & 1.23\%   & 16.68\%  & 0.96\%  \\
                                  & APL & 16.09\%  & 0.53\% & 19.84\%  & 0.75\%   & 26.45\%  & 2.71\%  \\
                                  & MEM & 2.92\%   & 0.65\% & 4.89\%   & 0.61\%   & 3.73\%   & 1.72\%  \\ \hline
\multirow{3}{*}{Petclinic}        & DEV & 12.49\%  & 1.97\% & 3.07\%   & 3.29\%   & -0.16\%  & 3.02\%  \\
                                  & APL & 4.48\%   & 3.61\% & 1.80\%   & 1.45\%   & -1.49\%  & 4.75\%  \\
                                  & MEM & 3.48\%   & 1.41\% & 1.55\%   & 2.34\%   & -4.51\%  & 4.82\%  \\ \hline
\multirow{3}{*}{Shopizer}         & DEV & 10.88\%  & 0.55\% & 14.06\%  & 1.70\%   & 0.95\%   & 0.78\%  \\
                                  & APL & 7.76\%   & 1.54\% & 5.93\%   & 2.23\%   & -16.82\% & 1.01\%  \\
                                  & MEM & 25.65\%  & 1.43\% & 19.56\%  & 1.85\%   & 19.90\%  & 3.01\%  \\ \hline
\multirow{3}{*}{Thingsboard}      & DEV & 0.26\%   & 0.12\% & -0.05\%  & 0.42\%   & -0.53\%  & 4.25\%  \\
                                  & APL & 0.10\%   & 1.58\% & -0.55\%  & 2.27\%   & 1.65\%   & 5.25\%  \\
                                  & MEM & 0.10\%   & 0.46\% & -0.71\%  & 0.96\%   & -1.51\%  & 3.09\%  \\ \midrule
\multirow{3}{*}{\textbf{Average}} & DEV & 8.68\%   & 0.45\% & 9.17\%   & 13.38\%  & 18.79\%  & 9.10\%  \\
                                  & APL & 1.68\%   & 0.87\% & 19.21\%  & 1.18\%   & 20.24\%  & 8.25\%  \\
                                  & MEM & 4.00\%   & 0.57\% & 3.49\%   & 0.48\%   & 3.35\%   & 4.99\%  \\ \bottomrule
\end{tabular}%
}
\end{table}
	
\begin{figure}
    \centering
    \includegraphics[scale=0.75,angle=90]{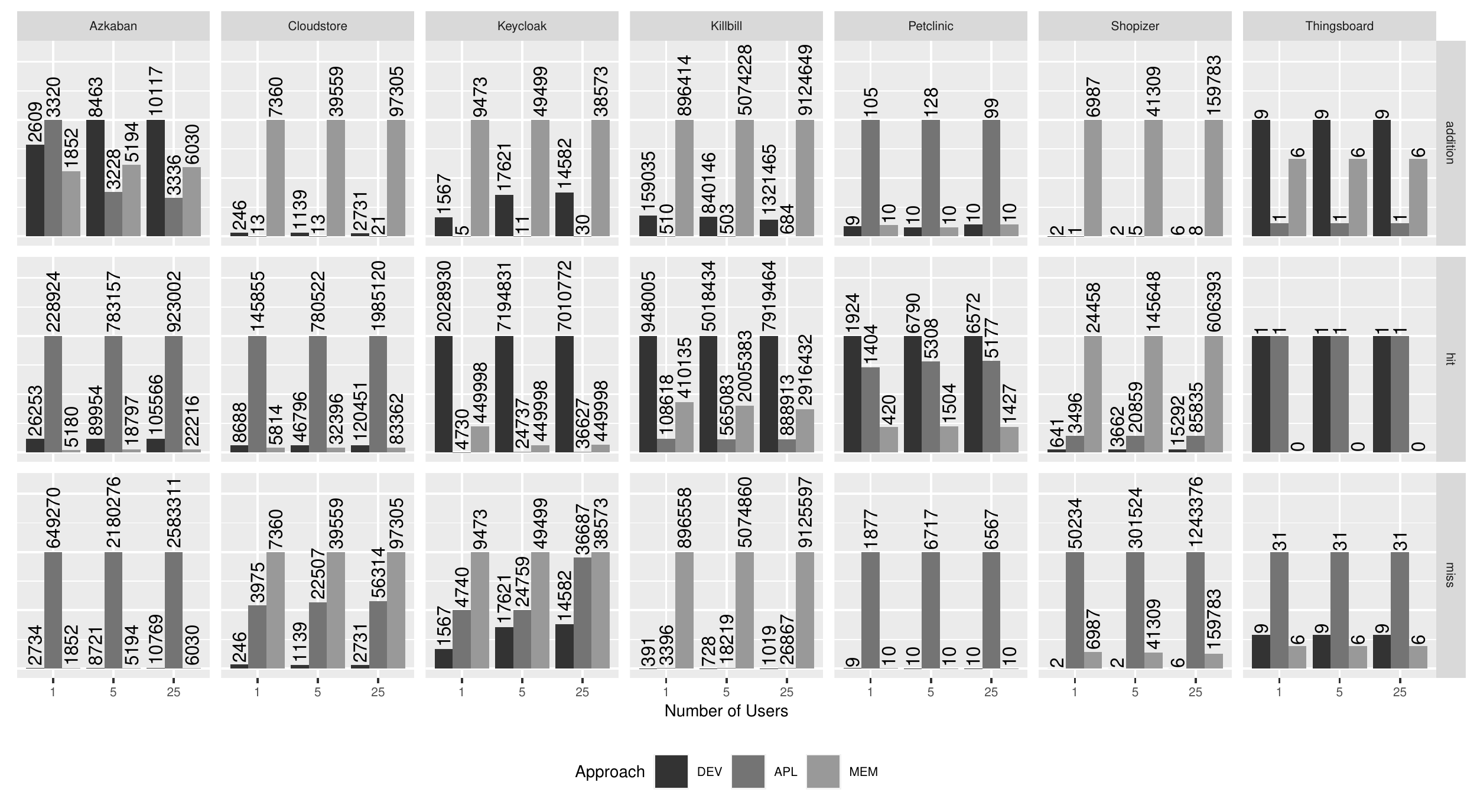}
    \caption{Normalised additions, hits and misses by approach for each application.}
    \label{figure:cache-balance}
\end{figure}

In addition to the throughput, we examine performance metrics of the cache, measuring hits and misses, as shown in Figure~\ref{figure:cache-balance} and Table~\ref{table:cache-balance}. As above, hits and misses are normalised in the chart with the actual numbers informed in the labels. We also present the addition of entries in the cache. This is helpful to understand the APL results because this approach caches only particular method inputs. Consequently, when there is a cache miss, the calculated result is not necessarily cached. For the MEM version, in turn, the number of additions is equal to the number of misses. Divergences between additions and misses in the DEV version are due to the manual update of cache entries that had their data source changed (additions that are not a consequence of a miss).

\begin{table*}
\renewcommand{\arraystretch}{1.1}
\renewcommand{\tabcolsep}{0.15cm}
\caption{Additions, Hits and Misses by Application.}
\label{table:cache-balance}
\centering
\resizebox*{!}{0.9\textheight}{%
\begin{tabular}{@{}rclr | rrr | rrr | rrr@{}}
\toprule
\multicolumn{1}{c}{\multirow{2}{*}{}} &
\multirow{2}{*}{} &
\multicolumn{1}{c}{\multirow{2}{*}{}} &
\multicolumn{1}{c|}{\multirow{2}{*}{}} &
\multicolumn{3}{c|}{\textbf{Additions}} &
\multicolumn{3}{c|}{\textbf{Hits}} & 
\multicolumn{3}{c}{\textbf{Misses}} \\
\multicolumn{1}{c}{\rotatebox[origin=l]{90}{\textbf{Project}}} &
\multicolumn{1}{c}{\rotatebox[origin=l]{90}{\textbf{Users}}} &
\multicolumn{1}{c}{\rotatebox[origin=l]{90}{\textbf{Approach}}} &
\multicolumn{1}{c|}{\rotatebox[origin=l]{90}{\textbf{Throughput}}} &
\multicolumn{1}{c}{\rotatebox[origin=l]{0}{\textbf{Median}}} &
\multicolumn{1}{c}{\rotatebox[origin=l]{0}{\textbf{Mean}}} &
\multicolumn{1}{c|}{\rotatebox[origin=l]{0}{\textbf{St. Dev.}}} &
\multicolumn{1}{c}{\rotatebox[origin=l]{0}{\textbf{Median}}} &
\multicolumn{1}{c}{\rotatebox[origin=l]{0}{\textbf{Mean}}} &
\multicolumn{1}{c|}{\rotatebox[origin=l]{0}{\textbf{St. Dev.}}} &
\multicolumn{1}{c}{\rotatebox[origin=l]{0}{\textbf{Median}}} &
\multicolumn{1}{c}{\rotatebox[origin=l]{0}{\textbf{Mean}}} &
\multicolumn{1}{c}{\rotatebox[origin=l]{0}{\textbf{St. Dev.}}} \\ \midrule
\multirow{9}{*}{\rotatebox[origin=c]{90}{Azkaban}}           & \multirow{3}{*}{1}           & DEV          & 9.94\%              & 1,304.00       & 869.67         & 752.29         & 26,253.00      & 26,253.00       & 0.00             & 1,367.00      & 1,367.00       & 1,931.82         \\
                                   &                              & APL          & -13.35\%            & 665.00         & 664.00         & 27.96          & 12,512.00      & 76,308.00       & 117,029.39       & 665.00        & 108,211.67     & 261,269.87       \\
                                   &                              & MEM          & 2.40\%              & 926.00         & 926.00         & 552.96         & 2,590.00       & 2,590.00        & 3,587.86         & 926.00        & 926.00         & 552.96           \\ \cmidrule(lr){5-13}
                                   & \multirow{3}{*}{5}           & DEV          & 1.36\%              & 4,231.00       & 2,821.00       & 2,442.19       & 89,954.00      & 89,954.00       & 0.00             & 4,360.50      & 4,360.50       & 6,165.26         \\
                                   &                              & APL          & -1.08\%             & 604.00         & 645.60         & 63.92          & 44,148.00      & 261,052.33      & 398,027.51       & 635.50        & 363,379.33     & 881,140.06       \\
                                   &                              & MEM          & 0.51\%              & 2,597.00       & 2,597.00       & 2,798.73       & 9,398.50       & 9,398.50        & 13,069.45        & 2,597.00      & 2,597.00       & 2,798.73         \\ \cmidrule(lr){5-13}
                                   & \multirow{3}{*}{25}          & DEV          & 0.67\%              & 5,058.00       & 3,372.33       & 2,919.66       & 105,566.00     & 105,566.00      & 0.00             & 5,384.50      & 5,384.50       & 7,613.42         \\
                                   &                              & APL          & 0.42\%              & 594.00         & 667.20         & 100.95         & 52,385.00      & 307,667.33      & 468,775.34       & 677.00        & 430,551.83     & 1,044,222.15     \\
                                   &                              & MEM          & 0.08\%              & 3,015.00       & 3,015.00       & 3,245.62       & 11,108.00      & 11,108.00       & 15,375.33        & 3,015.00      & 3,015.00       & 3,245.62         \\ \midrule
\multirow{9}{*}{\rotatebox[origin=c]{90}{Cloudstore}}        & \multirow{3}{*}{1}           & DEV          & 6.39\%              & 110.00         & 82.00          & 52.00          & 2,310.00       & 2,896.00        & 1,114.28         & 110.00        & 82.00          & 52.00            \\
                                   &                              & APL          & -10.59\%            & 1.00           & 4.33           & 5.77           & 72,927.50      & 72,927.50       & 92,511.49        & 1,916.00      & 1,325.00       & 1,139.87         \\
                                   &                              & MEM          & -7.29\%             & 3,680.00       & 3,680.00       & 93.34          & 5,814.00       & 5,814.00        & 0.00             & 3,680.00      & 3,680.00       & 93.34            \\ \cmidrule(lr){5-13}
                                   & \multirow{3}{*}{5}           & DEV          & 5.43\%              & 550.00         & 379.67         & 309.87         & 12,547.00      & 15,598.67       & 5,463.26         & 550.00        & 379.67         & 309.87           \\
                                   &                              & APL          & -2.83\%             & 1.00           & 4.33           & 5.77           & 390,261.00     & 390,261.00      & 494,035.71       & 10,914.00     & 7,502.33       & 6,496.28         \\
                                   &                              & MEM          & -1.07\%             & 19,779.50      & 19,779.50      & 468.81         & 16,198.00      & 16,198.00       & 22,900.36        & 19,779.50     & 19,779.50      & 468.81           \\ \cmidrule(lr){5-13}
                                   & \multirow{3}{*}{25}          & DEV          & 10.52\%             & 1,335.00       & 910.33         & 767.81         & 32,045.00      & 40,150.33       & 14,143.77        & 1,335.00      & 910.33         & 767.81           \\
                                   &                              & APL          & -0.33\%             & 5.00           & 7.00           & 3.46           & 992,560.00     & 992,560.00      & 1,256,818.66     & 27,308.00     & 18,771.33      & 16,268.81        \\
                                   &                              & MEM          & -1.31\%             & 48,652.50      & 48,652.50      & 1,181.58       & 41,681.00      & 41,681.00       & 58,923.21        & 48,652.50     & 48,652.50      & 1,181.58         \\ \midrule
\multirow{9}{*}{\rotatebox[origin=c]{90}{Keycloak}}          & \multirow{3}{*}{1}           & DEV          & 9.56\%              & 6.00           & 223.86         & 571.12         & 53,938.00      & 289,847.14      & 418,788.59       & 6.00          & 223.86         & 571.12           \\
                                   &                              & APL          & 7.32\%              & 5.00           & 5.00           & 0.00           & 4,730.00       & 4,730.00        & 0.00             & 2,370.00      & 2,370.00       & 3,344.62         \\
                                   &                              & MEM          & 0.75\%              & 1.00           & 1,894.60       & 2,592.92       & 49,999.00      & 149,999.33      & 173,205.66       & 1.00          & 1,894.60       & 2,592.92         \\ \cmidrule(lr){5-13}
                                   & \multirow{3}{*}{5}           & DEV          & 27.21\%             & 5.00           & 2,517.29       & 6,638.07       & 182,420.00     & 1,027,833.00    & 1,429,579.42     & 5.00          & 2,517.29       & 6,638.07         \\
                                   &                              & APL          & 111.34\%            & 11.00          & 11.00          & 0.00           & 24,737.00      & 24,737.00       & 0.00             & 12,379.50     & 12,379.50      & 17,491.70        \\
                                   &                              & MEM          & -0.33\%             & 1.00           & 9,899.80       & 13,554.49      & 49,999.00      & 149,999.33      & 173,205.66       & 1.00          & 9,899.80       & 13,554.49        \\ \cmidrule(lr){5-13}
                                   & \multirow{3}{*}{25}          & DEV          & 103.40\%            & 3.00           & 2,083.14       & 5,494.73       & 164,842.00     & 1,001,538.86    & 1,371,929.40     & 3.00          & 2,083.14       & 5,494.73         \\
                                   &                              & APL          & 131.82\%            & 30.00          & 30.00          & 0.00           & 36,627.00      & 36,627.00       & 0.00             & 18,343.50     & 18,343.50      & 25,899.20        \\
                                   &                              & MEM          & 7.04\%              & 1.00           & 7,714.60       & 10,562.28      & 49,999.00      & 149,999.33      & 173,205.66       & 1.00          & 7,714.60       & 10,562.28        \\ \midrule
\multirow{9}{*}{\rotatebox[origin=c]{90}{Killbill}}          & \multirow{3}{*}{1}           & DEV          & 11.23\%             & 50.50          & 15,903.50      & 34,280.24      & 4,239.50       & 94,800.50       & 194,375.53       & 25.50         & 48.88          & 51.66            \\
                                   &                              & APL          & 16.08\%             & 22.00          & 26.84          & 12.04          & 250.00         & 18,103.00       & 37,541.39        & 23.00         & 178.74         & 458.68           \\
                                   &                              & MEM          & 2.92\%              & 46,036.00      & 37,350.58      & 34,451.40      & 5,488.00       & 58,590.71       & 77,149.21        & 46,036.00     & 37,356.58      & 34,445.19        \\ \cmidrule(lr){5-13}
                                   & \multirow{3}{*}{5}           & DEV          & 13.08\%             & 149.00         & 84,014.60      & 181,845.49     & 14,689.50      & 501,843.40      & 1,033,550.66     & 76.00         & 91.00          & 78.52            \\
                                   &                              & APL          & 19.84\%             & 23.00          & 26.47          & 16.11          & 46.00          & 62,787.00       & 163,814.67       & 24.00         & 958.89         & 3,123.92         \\
                                   &                              & MEM          & 4.89\%              & 240,052.50     & 211,426.17     & 210,213.97     & 482.00         & 100,269.15      & 262,508.42       & 240,052.50    & 211,452.50     & 210,186.83       \\ \cmidrule(lr){5-13}
                                   & \multirow{3}{*}{25}          & DEV          & 16.68\%             & 204.00         & 132,146.50     & 286,228.24     & 21,123.00      & 791,946.40      & 1,633,101.93     & 124.00        & 127.38         & 116.28           \\
                                   &                              & APL          & 26.45\%             & 26.00          & 36.00          & 21.87          & 7.00           & 68,377.92       & 217,297.47       & 29.00         & 1,414.05       & 4,608.72         \\
                                   &                              & MEM          & 3.73\%              & 379,123.00     & 380,193.71     & 413,077.32     & 2,671.50       & 145,821.60      & 409,644.16       & 379,123.00    & 380,233.21     & 413,039.81       \\ \midrule
\multirow{9}{*}{\rotatebox[origin=c]{90}{Petclinic}}         & \multirow{3}{*}{1}           & DEV          & 12.49\%             & 9.00           & 9.00           & 0.00           & 1,924.00       & 1,924.00        & 0.00             & 9.00          & 9.00           & 0.00             \\
                                   &                              & APL          & 4.48\%              & 10.00          & 35.00          & 44.17          & 420.00         & 468.00          & 414.09           & 10.00         & 625.67         & 1,067.23         \\
                                   &                              & MEM          & 3.48\%              & 10.00          & 10.00          & 0.00           & 420.00         & 420.00          & 0.00             & 10.00         & 10.00          & 0.00             \\ \cmidrule(lr){5-13}
                                   & \multirow{3}{*}{5}           & DEV          & 3.07\%              & 10.00          & 10.00          & 0.00           & 6,790.00       & 6,790.00        & 0.00             & 10.00         & 10.00          & 0.00             \\
                                   &                              & APL          & 1.80\%              & 10.00          & 42.67          & 56.58          & 1,504.00       & 1,769.33        & 1,671.87         & 10.00         & 2,239.00       & 3,860.74         \\
                                   &                              & MEM          & 1.55\%              & 10.00          & 10.00          & 0.00           & 1,504.00       & 1,504.00        & 0.00             & 10.00         & 10.00          & 0.00             \\ \cmidrule(lr){5-13}
                                   & \multirow{3}{*}{25}          & DEV          & -0.16\%             & 10.00          & 10.00          & 0.00           & 6,572.00       & 6,572.00        & 0.00             & 10.00         & 10.00          & 0.00             \\
                                   &                              & APL          & -1.49\%             & 8.00           & 33.00          & 44.17          & 1,429.00       & 1,725.67        & 1,648.15         & 8.00          & 2,189.00       & 3,778.47         \\
                                   &                              & MEM          & -4.51\%             & 10.00          & 10.00          & 0.00           & 1,427.00       & 1,427.00        & 0.00             & 10.00         & 10.00          & 0.00             \\ \midrule
\multirow{9}{*}{\rotatebox[origin=c]{90}{Shopizer}}          & \multirow{3}{*}{1}           & DEV          & 10.88\%             & 1.00           & 1.00           & 0.00           & 320.50         & 320.50          & 280.72           & 1.00          & 1.00           & 0.00             \\
                                   &                              & APL          & 7.76\%              & 1.00           & 1.00           & 0.00           & 3,496.00       & 3,496.00        & 0.00             & 4,672.00      & 4,186.17       & 1,359.45         \\
                                   &                              & MEM          & 25.65\%             & 1.00           & 317.59         & 687.42         & 786.50         & 2,038.17        & 2,161.66         & 1.00          & 317.59         & 687.42           \\ \cmidrule(lr){5-13}
                                   & \multirow{3}{*}{5}           & DEV          & 14.06\%             & 1.00           & 1.00           & 0.00           & 1,831.00       & 1,831.00        & 1,673.01         & 1.00          & 1.00           & 0.00             \\
                                   &                              & APL          & 5.93\%              & 5.00           & 5.00           & 0.00           & 20,859.00      & 20,859.00       & 0.00             & 28,123.00     & 25,127.00      & 8,225.69         \\
                                   &                              & MEM          & 19.56\%             & 5.00           & 1,877.68       & 3,946.12       & 1,660.00       & 11,203.69       & 12,975.28        & 5.00          & 1,877.68       & 3,946.12         \\ \cmidrule(lr){5-13}
                                   & \multirow{3}{*}{25}          & DEV          & 0.95\%              & 3.00           & 3.00           & 0.00           & 7,646.00       & 7,646.00        & 7,085.21         & 3.00          & 3.00           & 0.00             \\
                                   &                              & APL          & -16.82\%            & 8.00           & 8.00           & 0.00           & 85,835.00      & 85,835.00       & 0.00             & 115,821.00    & 103,614.67     & 33,797.44        \\
                                   &                              & MEM          & 19.90\%             & 8.00           & 7,262.86       & 14,563.39      & 6,741.00       & 37,899.56       & 51,136.50        & 8.00          & 7,262.86       & 14,563.39        \\ \midrule
\multirow{9}{*}{\rotatebox[origin=c]{90}{Thingsboard}}       & \multirow{3}{*}{1}           & DEV          & 0.26\%              & 9.00           & 9.00           & 0.00           & 1.00           & 1.00            & 0.00             & 9.00          & 9.00           & 0.00             \\
                                   &                              & APL          & 0.10\%              & 1.00           & 1.00           & 0.00           & 1.00           & 1.00            & 0.00             & 31.00         & 31.00          & 0.00             \\
                                   &                              & MEM          & 0.10\%              & 3.00           & 3.00           & 0.00           & 0.00           & 0.00            & 0.00             & 3.00          & 3.00           & 0.00             \\ \cmidrule(l){5-13} 
                                   & \multirow{3}{*}{5}           & DEV          & -0.05\%             & 9.00           & 9.00           & 0.00           & 1.00           & 1.00            & 0.00             & 9.00          & 9.00           & 0.00             \\
                                   &                              & APL          & -0.55\%             & 1.00           & 1.00           & 0.00           & 1.00           & 1.00            & 0.00             & 31.00         & 31.00          & 0.00             \\
                                   &                              & MEM          & -0.71\%             & 3.00           & 3.00           & 0.00           & 0.00           & 0.00            & 0.00             & 3.00          & 3.00           & 0.00             \\ \cmidrule(lr){5-13}
                                   & \multirow{3}{*}{25}          & DEV          & -0.53\%             & 9.00           & 9.00           & 0.00           & 1.00           & 1.00            & 0.00             & 9.00          & 9.00           & 0.00             \\
                                   &                              & APL          & 1.65\%              & 1.00           & 1.00           & 0.00           & 1.00           & 1.00            & 0.00             & 31.00         & 31.00          & 0.00             \\
                                   &                              & MEM          & -1.51\%             & 3.00           & 3.00           & 0.00           & 0.00           & 0.00            & 0.00             & 3.00          & 3.00           & 0.00             \\ \bottomrule
\end{tabular}
}
\end{table*}

The results of the compared approach present diverging results across the various applications as well as the different number of users within a same application. APL has a higher variance in the relative throughput (\textbf{Evidence 7}). It ranges from -16.82\% (Shopizer, 25 users) to 131.82\% (Keycloak, 25 users), while the relative throughput obtained with MEM varies from -7.29\% (Cloudstore, 1 user) to 25.65\% (Shopizer, 1 user). Consequently, it is not possible to claim that any of the approaches (APL, MEM or DEV) leads to a higher performance or higher hit-ratio. To understand why this happened, we next investigate in depth the issues that occurred with the different approaches that led to these results.

We begin by looking at the results obtained with APL. As discussed, this approach selects particular inputs of cacheable methods to be cached, and its recommendations often included a single method input to be cached. Caching a pair of method inputs and output that is not frequent uses space in the cache without providing any gain. However, the choice for particular inputs to be cached causes performance decays due to two reasons. (1) As shown in Table~\ref{table:inputs-discarded}, there are many recurrent inputs in cached methods. This means that inputs, possibly not observed during the period to generate recommendations, could provide benefits if cached. This situation can be seen in the Shopizer application, which has a high number of misses but a low number of additions. It also has a low number of distinct inputs that are not cached. (2) Even in the cases where the not cached inputs are not frequent, if there is a wide range of inputs, the cost of managing the cache within a method might not compensate the gains provided by caching a particular frequent input. This is the case of the Azkaban application, which also has a high number of misses and a low number of additions, but the number of distinct inputs that are not cached is higher. In both cases, APL achieved the worst results in most of the user configurations. Therefore, it is important to not only search for inputs that would provide gains if cached (because of its frequency or time to compute), but also analyse the range of possible inputs (\textbf{Evidence 8}). Furthermore, it is important to distinguish when a cacheable method receives a particular frequent input and when a method tends to receive frequent inputs, which change over time.

While APL focuses not only on identifying methods that are feasible to be cached but also a subset that would provide gains if cached, MEM aims to identify all methods that can be cached (discarding those that are cheap to compute based on a fixed threshold). This justifies why MEM not always provides performance improvements---some of the recommendations, although cacheable (i.e.\ the same inputs produce the same output), do not provide gains if cached because it is not too frequent or expensive to compute. MEM assumes that developers decide which recommendations to cache. To ease this analysis, the approach ranks the recommendations by calculating the time saved by caching a method based on the potential hits and its computation time. We observed that highly ranked recommendations in fact are those that lead to more hits during the execution  (\textbf{Evidence 9}). A reason for MEM to achieve low throughput in some cases, such as for Cloudstore, is that most of the recommendations indicated that methods should be cached using a single-instance cache. This caused the cache entry associated with a method to be constantly replaced with a new pair of inputs-output, without reusing the previous one (i.e.\ cache thrashing) (\textbf{Evidence 10}). This led to a high number of misses with a low number of hits.

We used as much as possible of the original implementations of APL and MEM and, in the cases when it was not possible, we used the original implementation as a basis. When analysing our results, we observed that implementation details also largely contribute to the obtained results. MEM compares if objects are equal with the \texttt{equals()} method, while APL serialises objects. This is because APL keeps a table at runtime with the specific inputs of a method that should be cached. On the one hand, serialising and deserialising objects at runtime have a cost, resulting in a penalty in the gained performance. On the other hand, some of the compared objects had properties that are not included in the serialisation. These, if taken into account, could wrongly lead to the conclusion that two objects are not equal when they are. Consequently, APL could reuse methods that MEM considered not cacheable by mistakenly identifying that two objects are different.

Finally, we emphasise that the results obtained with the developers' cached methods are presented as a baseline. As the workload used by the developers to make caching decisions is not available and is different from ours, their decisions can be not optimal in our scenario. Nevertheless, the version of the application that included methods cached by developers (DEV) still obtained the best results for the majority of the applications. This provides evidence of the adequacy of the workload used for this study.


\section{Discussion}
\label{section:discussion}

Based on the presented results, we now discuss implications from our empirical evaluation, which give directions for future application-level approaches. We also examine the threats to the validity of our study.

\subsection{Causes for Missed Opportunities}

APL and MEM provided caching recommendations that result in performance gains. However, by analysing the caching opportunities that they \emph{missed}, we observed that both APL and MEM do not take into account other factors that are fundamental to provide better results in terms of throughput and hits. These factors are listed as follows.

\begin{itemize}

	\item \textbf{Time-to-live (TTL}). When putting elements in the cache, we used the TTL used by developers in the target application. However, the TTL plays a key role in the cache management. If a cache entry expires before it is reused, it provide no gains. None of the approaches analysed the distribution of repeated calls and the selection of TTL values. They only take into account the complete set of traces regardless of how close or spread the repeated calls are.
	
	\item \textbf{Changeability}. A cacheable method is feasible to be cached even if the same inputs can produce a different output. In this case, cache entries can be evicted or stale data are allowed for short periods of time. If this is taken into consideration, more methods that could provide performance gains could be cached. MEM discards them. APL, in turn, considers this but, because it does not recommend a TTL to reduce the chances of stale data, it could led to bugs if it is used to automatically manage the application-level caching.
	
	\item \textbf{Shareability}. APL achieved better results for many users in most of the cases. This is mainly because, to select methods to cache, it considers if a certain computation is reused for different users (better results were not achieved with Shopizer due to the issue of caching only a particular input of a method). This is an important aspect that is not considered in MEM.
	
	\item \textbf{Prediction of Hits}. Both approaches analyse a given set of traces to suggest recommendations. They assume that estimated hits in the future are equal to the hits that would occur with the given set of traces. This occurs mainly in APL because it assumes that a method with frequent inputs in the analysed traces would receive calls with the same input in the future. Consequently, it is important for the approaches to verify if they can \emph{predict} hits and performance gains, not only assume that the given set of traces is completely representative of future workloads. 
	
\end{itemize}

\subsection{Recommendations for Future Approaches}

The results do not indicate that there is a best approach to recommend methods to cache. However, they revealed many circumstances in which each approach are unable to provide adequate recommendations, which are the basis for developing future application-level caching supporting tools. Key circumstances are highlighted throughout the previous section as \emph{pieces of evidence}. These evidences are related to adjustments that must be done in the evaluated approaches. Consequently, based on our experiment, we derived various lessons learned, which are categorised into seven groups, discussed as follows.

\begin{enumerate}

	\item \textbf{Explore method changeability and time-to-live (TTL) values (\textbf{Evidence 7}).} MEM considers methods as cacheable when a method \emph{always} produce a same output for a given input. This significantly limits the choice for methods to cache. APL, in contrast, admits methods that have low changeability, that is, methods that in general have this behaviour but not always. Developers also make choices as APL. However, they control data freshness by choosing adequate TTL values or doing manual cache content invalidation. Consequently, it is crucial for approaches to further explore this issue and explore how to choose adequate TTLs for particular methods. Moreover, the choice for TTL also allows methods that have many calls in short periods of time (but less than other methods considering a larger time frame) to be cached making an effective use of the cache by being cached with short TTL values. 

	\item \textbf{Inspect the domain of method parameters (\textbf{Evidence 5 and 11})}.  APL concerns frequent and/or expensive method inputs. Although some method inputs promote benefits if cached, the costs of managing the cache to cache one or a few inputs when a method is called a much higher number of times with other inputs can result in a performance decrease. Therefore, it is fundamental not only to search for frequent/expensive inputs, but also to look at the range of values provided as parameters. MEM take a step towards this direction by recommending the size of the cache, but the simplistic idea of having a single instance cache (i.e.\ one entry only) can result in cache thrashing.

	\item \textbf{Consider the method body to detect invalid recommendations. (\textbf{Evidence 1 and 2}).} Both investigated approaches recommended many invalid recommendations, because they inspect solely the values of method inputs and outputs. This prevents the complete automation of application-level caching because, for various methods, there are internal operations that cannot be skipped. As a consequence, it is important for approaches to examine internal method operations to discard such methods. This would largely reduce the developers' effort, who currently manually perform this task.

	\item \textbf{Investigate the trade-off between caching coarse- or fine-grained methods (\textbf{Evidence 4}).}  Our study showed that in some situations developers cached fine-grained methods (a callee method) while APL and MEM recommended caching coarse-grained methods (a caller method). Caching the callee promotes a higher number of hits (because it may be called by multiple methods), while caching the caller method saves more computation time (because it reuses more operations). The trade-off between choosing a callee or caller method should be further investigated, as in different situations a different choice is optimal.

	\item \textbf{Consider the identification of methods feasible to cache and, from those, the selection of methods that improve performance as separate tasks (\textbf{Evidence 1--3 and 9}).}  APL not only aims to identify methods that are feasible to be cached, but also to select a subset of methods that promotes more benefits if cached. Differently, MEM focuses on identifying methods that are feasible to be cached (using a more strict definition) and ranks these methods. However, both approaches still recommend many invalid recommendations. Therefore, the automation of application-level caching could be split into two separate tasks: (i) the recommendation of methods that are feasible to be cached, that is, those do not cause bugs if cached; and (ii) the recommendation of a subset of methods feasible to be cached that would promote performance gains if cached. By guaranteeing that the recommendations of (ii) do not cause bugs if cached, it is possible to automate it using an adaptive approach. Because the application workload varies at runtime, caching decisions must be constantly revised. This automation relieves developers from a time-consuming task. 
	
As discussed by the authors of MemoizeIt, detecting whether a recommendation is completely safe to cache is a costly operation. Detecting this requires recording and comparing the complete previous state of the application before and after caching, including what is outside the boundaries of the application (such as in a database). In order to reduce such a cost, future approaches could perform an assisted static analysis (offline or at compilation time) to mark methods as uncacheable. Marking such methods would reduce the costs of the analysis by excluding all methods that contain any write operation in static variables, inputs received in the method, or even in the outside environment. With developers in the loop, next generation approaches could make safer and more reliable recommendations.

	\item \textbf{Provision of Explanations (\textbf{Evidence 5 and 8}).} There are many aspects that should be considered to decide methods to cache (input-output behaviour, the domain of input values, the invocation frequency, the computation time, and so on). Given that the complete automation of application-level caching is still not possible due to invalid recommendations and the choice for methods that lead to performance decays, recommender approaches could provide complementary information to the recommendation so that developers could make more informed decisions of what methods to cache.

	\item \textbf{Consider implementation aspects (\textbf{Evidence 6 and 10}).} Our study has shown that how application-level caching is implemented has a high impact on the application performance. More specifically, although the serialisation of objects takes time to be performed, it led to different results when objects are compared. Thus, these implementation issues must be seriously taken into account because they impact on the performance in many ways.

\end{enumerate}

\subsection{Threats to Validity}

We next report how we addressed the identified threats to validity in our study.

\subsubsection{Construction Validity}

Our study required the implementation and adaptation of existing approaches as well as the target applications. To prevent the introduction of bugs in APL and MEM, we validated our implementation carefully analysing the behaviour and outputs of each developed module using a small-scale set of inputs. Moreover, we logged every critical operation in order to verify if the implementation followed the expected flow. We also used an existing cache component, which increases the reliability of our implementation, with respect to this aspect. Regarding the implementation of different versions of the applications, they were created by adding or removing localised cache snippets. This task was performed with a careful analysis and the execution of unit tests (if present).

The compared approaches use thresholds. Although we used the values provided by the authors of these approaches, we explored a range of values for the thresholds using testing data. The best results were those obtained with the default thresholds. We also had to choose other parameters, such as the amount of time to execute the applications and number of simultaneous users. Varying the amount of time both to generate traces and to collect measurements produced consistent results. The same occurred with other numbers of simultaneous users.


To avoid the impact of external factors on the measured throughput, such as operating system tasks, we executed using docker images the same set of requests ten times and analysed the variance of results. Moreover, no other tasks were being carried out in the used machine. Even the simulation of requests was performed in a different machine.

\subsubsection{External Validity}

A threat to the external validity of our study is our selected applications, which one can argue that are not representative. In order to address this, we followed a systematic procedure to select GitHub repositories, which include applications of different domains and sizes. Another threat is the generated workloads and data stored in the databases, which could be not representative of the applications in real settings. Note that both APL and MEM generate recommendations for a specific workload and data. Consequently, our generated synthetic workloads and data affected APL and MEM equally. Moreover, we extracted a navigation graph according to the organisation of pages within each web application to generate realistic sequences of requests, avoiding making particular methods more frequent than they would be in a real scenario. We also added a probability of a user closing the browser session while accessing the web application to make the interaction even more realistic.


An issue that may be seen as a threat is the marginal improvements in the throughput values provided by caching of both compared approaches as well as the developers' version. With respect to the differences of APL and MEM in comparison with the applications with no caching, as discussed, both approaches make some recommendations that provide performance improvements (adequate recommendations) and some that not only do not improve performance but also degrades it (inadequate recommendations). Consequently, the inadequate recommendations, instead of improving the performance, increase the computation time by accessing and adding  cached content. This justifies the variances in the results. With respect to the developers' version, we highlight that it was added as a reference. However, the workloads and data used for the applications are not the same that the developers used to make decisions of what to cache. Therefore, their decisions might not be optimal considering our simulated scenarios.


\section{Conclusion}
\label{section:conclusions}

The current demands of high performance and scalability in software applications lead developers to explore a wide range of techniques to meet these requirements. Application-level caching is an approach that has been increasingly being adopted. Although it addresses performance issues by reusing frequent and/or expensive computations, it comes with an effort to identify methods to cache and revise caching decisions periodically. Approaches have been proposed to support developers in this time-consuming task.

From the existing application-level caching approaches, two are able to identify and recommend cacheable methods. They have been individually evaluated with a small subset of applications. In this paper, we presented the results of a study in which we evaluated these approaches, namely APLCache and MemoizeIt, using a set of seven open-source web applications. As an evaluation in this direction is a challenge because involves many design decisions and parameters, we proposed a protocol to evaluate these approaches, with the aim of reducing possible bias in the evaluation. We adapted the original web applications, creating four versions of each: an uncached version, and three versions with cached methods selected by developers, APLCache and MemoizeIt. The results were analysed from two perspectives, which are the caching choices and performance (throughput, hits and misses). The obtained results revealed that both approaches are able to identify cacheable methods that improve performance if cached. However, their recommendations include many invalid recommendations as well as methods that do not promote expected benefits. We made an in-depth analysis of good and bad recommendations, which allowed us to derive seven lessons learned that pave the way for developing the next generation of application-level caching supporting approaches.

As future work, we aim to explore the raised issues unaddressed by APLCache and MemoizeIt. Our ongoing work is to exploit TTL values to identify better caching opportunities as well as to make an effective use of the cache, thus reducing the need for large cache components.

\begin{acknowledgements}
The authors would like to thank for CNPq grants ref. 131271/2018-0, ref. 313357/2018-8, and ref. 428157/2018-1. This study was financed in part by the Coordena\c{c}\~{a}o de Aperfei\c{c}oamento de Pessoal de N\'{i}vel Superior - Brasil (CAPES) - Finance Code 001.
\end{acknowledgements}

\appendix

\section{Reproducibility}
\label{appendix:reproducibility}

To reproduce and/or change the parameters of our experiment, it is required to have two computers with sufficient resources, both running Linux-based systems and having Java 11+, Maven, Docker CE and docker-compose installed. In our repository, we provide a \textbf{configure} file that helps to install such dependencies. Also, this configuration script automates the cloning of all the repositories of each application version in the proper folder, as well as the tools that we implemented and rely on. It is also possible to run our experiment in standalone mode (with only one computer). However, it may not generate reliable results, due to the interference of the requester application that would compete on resources with the running application. Finally, compiling and installing our tools through Maven is also required, which is automatically made by a script named \textbf{compile.sh}.

To the first phase of the experiment, we automatically detect all the applications (including new ones) within the folder that holds the NOCACHE version of applications through the script \textbf{trace.sh}. The hostname for the server machine must be informed. We then check whether the traces of the application were already collected in the outputs folder. If not yet executed, we send a signal to the server machine (which could also be the same machine) to bring the application up via a Docker container. It is expected to the server machine to be listening for requests with our Java tool called \textbf{RemoteExecutor}, as well as having all the mentioned configurations done. New applications are expected to have a \emph{Dockerfile}---as we did for our target applications---including its required steps to compile and to run, as well as a \emph{docker-compose} describing its dependencies, such as a database and its initialisation SQL. We also look for the files \emph{whitelist}, \emph{blacklist} and \emph{ignored} in order to configure which Java packages should be serialised by our \emph{JSONSerialiser} tool for the tracings. Also, including the dependency of our \emph{ApplicationTracer} in the \emph{pom} or \emph{gradle} file is required, so we can automatically inject code in every method of new applications. We then automatically wait for the application to start and once started we start firing requests. For new applications, describing the JSON file containing the workload graph as we did for our applications in the \emph{workloads} folder is required for this step. When finished with the tracing, we automatically tear the application down and clean its created files and data.

To the recommendation of methods based on the collected traces, it is just needed to execute the approaches using as input the trace file generated in the first phase. Manual analysis of the recommendations, as well as the manual implementation of the caching with our \emph{Cache} component, is required. Particularly for APL, the generated files containing the recommended inputs (which is already included in our repository) are required to be in the outputs folder. It is also important, for APL, that the same serialisation parameters adopted in the first phase are used in the second phase, so the inputs can match. For new applications, the Maven/Gradle dependency of our Cache component must be added. For new approaches, it is required that they are able to read and process the tracing file generated in the first phase so they can give their recommendations.

To the second phase of the experiment, we automatically detect all the applications and start firing the requests through our script named \textbf{run.sh}.The hostname for the server machine must be informed. If not yet generated the workloads, we automatically do it. We then start sampling all the ten executions for all the groups of simulated users collecting our metrics in the disk. Particular executions that already succeeded are skipped. It is expected in this phase that each application version is in the proper folder and includes the files, configurations and caching as mentioned before.

To aggregate the results produced in the outputs folder, the script \textbf{reduce.sh} helps on initialising the CSV files and executing our tools that calculate our metrics. The hostname for the server machine must be informed. In order to generate visualisations for the aggregated CSV files, the script \emph{plot.sh} under the analysis folder may be invoked.

%
%

\bibliographystyle{spbasic}      
\bibliography{references}   

\end{document}